\documentclass[preprint,aps,amsmath,amssymb,superscriptaddress,floatfix]{revtex4-2}

\usepackage[utf8]{inputenc}
\usepackage{setspace}
\usepackage{graphicx}
\usepackage{color}
\usepackage{soul}
\usepackage{hyperref}
\usepackage{miller}
\usepackage{pdfpages}
\usepackage{bbold}
\usepackage{gensymb}

\newcommand{\simlt}
      {\ifmmode       { \raisebox{-.8em}{$<$}\atop\sim}
         \else        {$\raisebox{-.8em}{$<$}\atop\sim$}
      \fi}

\makeatletter
\def\@hangfrom@section#1#2#3{\@hangfrom{#1#2}#3}
\def\@hangfroms@section#1#2{#1#2}
\makeatother

\makeatletter
\AtBeginDocument{\let\LS@rot\@undefined}
\makeatother

\usepackage[version=4]{mhchem}

\begin{document}


\title{Determining the superconducting order parameter of \ce{UPt3} using scanning tunneling microscopy}

\author{Rebecca Bisset}
\affiliation{SUPA, School of Physics and Astronomy, University of St Andrews, North Haugh, St Andrews, KY16 9SS, United Kingdom}
\author{Luke C. Rhodes}
\affiliation{SUPA, School of Physics and Astronomy, University of St Andrews, North Haugh, St Andrews, KY16 9SS, United Kingdom}
\author{Hugo Decitre}
\affiliation{SUPA, School of Physics and Astronomy, University of St Andrews, North Haugh, St Andrews, KY16 9SS, United Kingdom}
\author{Matthew J. Neat}
\affiliation{SUPA, School of Physics and Astronomy, University of St Andrews, North Haugh, St Andrews, KY16 9SS, United Kingdom}
\author{Ana Maldonado}
\affiliation{SUPA, School of Physics and Astronomy, University of St Andrews, North Haugh, St Andrews, KY16 9SS, United Kingdom}
\author{Andrew Huxley}
\affiliation{SUPA, School of Physics and Astronomy, University of Edinburgh, Kings Buildings, Edinburgh, EH9 3FD, United Kingdom}
\author{Carolina A. Marques}
\affiliation{SUPA, School of Physics and Astronomy, University of St Andrews, North Haugh, St Andrews, KY16 9SS, United Kingdom}
\author{Peter Wahl}
\email[Correspondence to: ]{wahl@st-andrews.ac.uk.}
\affiliation{SUPA, School of Physics and Astronomy, University of St Andrews, North Haugh, St Andrews, KY16 9SS, United Kingdom}
\affiliation{Physikalisches Institut, Universität Bonn, Nussallee 12, 53115 Bonn, Germany}

\date{\today}

\begin{abstract}
Superconductivity, a state in which electrical currents can flow without resistance, occurs because of pairing of electrons into quasiparticles with integer spin $S$. In practically all known superconducting materials, these pairs form a singlet with $S=0$. Finding a material that has triplet pairing, $S=1$, would have profound fundamental and technological implications. \ce{UPt3} has been a key candidate material for spin-triplet superconductivity. Because of a lack of direct evidence for the pairing symmetry, the nature of the superconducting pairing remains under debate. Here, we use ultra-low temperature scanning tunneling microscopy to resolve this question. Our data reveals a zero-bias Andreev bound state within the gap for a surface normal to the $c$-axis of \ce{UPt3}. The superconducting origin of the features is confirmed through vortex imaging. For triplet pairing, such an Andreev state is fragile against Rashba spin-splitting, whereas for singlet pairing it remains robust, classifying \ce{UPt3} as a spin-singlet superconductor with a chiral order parameter.
\end{abstract}


\maketitle
\section{Introduction}
Despite intense research efforts, the only superfluid that is so far confirmed as a condensate of spin-triplets is $^3$He, even though there is no fundamental reason why electrons should not form triplet pairs in superconductors. Uranium superconductors are a family of materials where there is promising evidence for unconventional pairing, and often even triplet pairing \cite{joynt_superconducting_2002,aoki_review_2019,aoki_unconventional_2022}, however the interpretation in terms of spin-singlet and -triplet is challenging because of the large non-negligible spin-orbit coupling in these materials. One of the uranium materials that shows strong evidence for spin-triplet superconductivity is \ce{UPt3}: a number of experiments have suggested that the superconducting state is time reversal symmetry breaking \cite{luke_muon_1993,schemm_observation_2014}, with the absence of a Knight shift on entering the superconducting state providing additional evidence \cite{tou_nonunitary_1998}. In addition, the superconducting phase diagram consists of multiple regions, seemingly all with different order parameters \cite{Adenwalla1990, Suderow1997, joynt_superconducting_2002} -- somewhat similar to the case of $^3$He, which exhibits two distinct superfluid phases.

\begin{figure}    \includegraphics[width=1\linewidth]{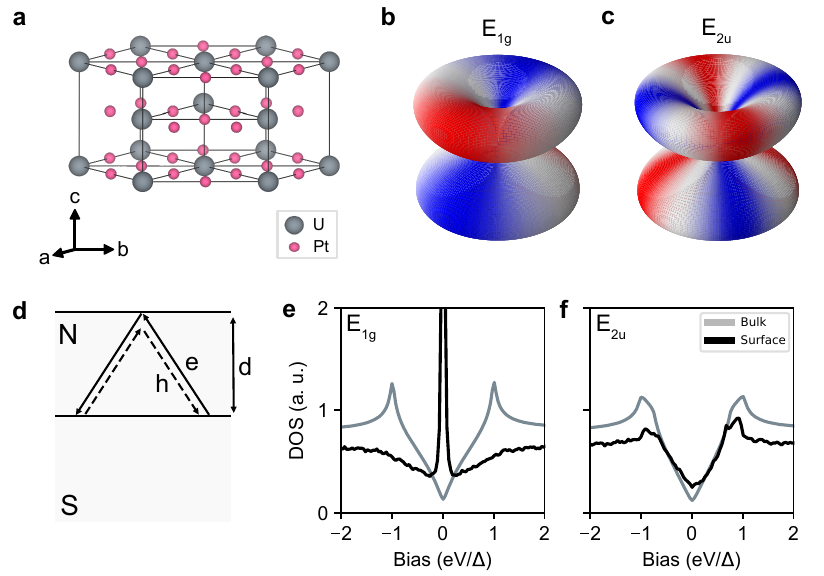}
    \caption{{\bf Superconducting order parameters of \ce{UPt3}.} (a) Crystal structure of \ce{UPt3}, showing planes of \ce{UPt3} stacked with a hexagonally-close-packed (hcp) alignment along the $c$-axis. (b, c) Graphical representations of the leading contenders for the order parameter in the low temperature, low field phase in \ce{UPt3}. (b) Spin-singlet order parameter with $E_{1g}$ symmetry, (c) spin-triplet order parameter with $E_{2u}$ symmetry. (d) Andreev reflection at a normal-superconducting interface, with electron paths indicated by solid arrows and hole paths shown by dashed arrows. The trajectories are specularly reflected at the vacuum interface, whereas at the $N-S$ interface, an electron with momentum $\mathbf{k}$ is reflected as a hole with momentum $-\mathbf{k}$ and vice-versa. (e, f) Local density of states (LDOS) at the surface of superconductors with singlet order parameter $E_{1g}$ and the triplet order parameter $E_{2u}$, respectively, including a non-negligible spin-orbit coupling (see suppl. sect.~\ref{andreevsection} for details).}
    \label{fig:intro_fig}
\end{figure}

The crystal structure of \ce{UPt3} has a hexagonal $D_{6h}$ point group symmetry, with planes of \ce{UPt3} stacked in a hexagonally close-packed arrangement (Fig.~\ref{fig:intro_fig}(a)). For the symmetry of the superconducting order parameter, two candidates have emerged as the leading contenders: a singlet order parameter with $E_{1g}$ symmetry (Fig.~\ref{fig:intro_fig}(b)) and a triplet order parameter with $E_{2u}$ symmetry (Fig.~\ref{fig:intro_fig}(c)) \cite{joynt_superconducting_2002}. Both are chiral order parameters with a nodal plane in the $a$-$b$ plane, and a point node along $c$, however one with a winding number $\nu=1$ ($E_{1g}$) and the other with a winding number $\nu=2$ ($E_{2u}$).

The tunneling spectrum at the surface can reveal key information about the symmetry of the order parameter through Andreev bound states \cite{hu_midgap_1994,Tanaka_Theory_1995,Honerkamp_Andreev_1998,christiansen_nodal_2025}. Andreev bound states were one of the earliest signatures of the $d$-wave nature of superconductivity in the cuprate superconductors \cite{deutscher_andreevsaint-james_2005}. The key criterion for the existence of surface Andreev bound states is a sign change of the order parameter across a plane parallel to the surface (Fig.~\ref{fig:intro_fig}(d)). For the leading contenders for the symmetry of the order parameter in \ce{UPt3}, $E_{1g}$ and $E_{2u}$, and a surface parallel to the basal plane, both result in a zero bias bound state. However for the $E_{2u}$ order parameter, the zero energy bound state is fragile and suppressed by Rashba spin-splitting at the surface due to spin-orbit coupling \cite{kobayashi_fragile_2015}. Because of the large spin-orbit coupling of uranium and platinum compounds, the zero bias bound states thus provide a clear discriminant: for a singlet order parameter with $E_{1g}$ symmetry a robust surface Andreev bound state at zero bias is expected (Fig.~\ref{fig:intro_fig}(e)), while for the triplet order parameter with $E_{2u}$ symmetry, the Andreev bound state is suppressed (Fig.~\ref{fig:intro_fig}(f)). 
A measurement of such a bound state therefore puts severe constraints on the symmetry of the order parameter. 

Here, we use scanning tunneling spectroscopy at ultra-low temperatures below $100\mathrm{mK}$ to determine the superconducting order parameter of \ce{UPt3} from the gap structure as detected in tunneling spectra. We show real space images of the vortex lattice, and extract the coherence length. Our results classify \ce{UPt3} as an unconventional superconductor with singlet pairing and an order parameter with $E_{1g}$ symmetry.

\section{Results}

\begin{figure}
    \includegraphics[width=1\linewidth]{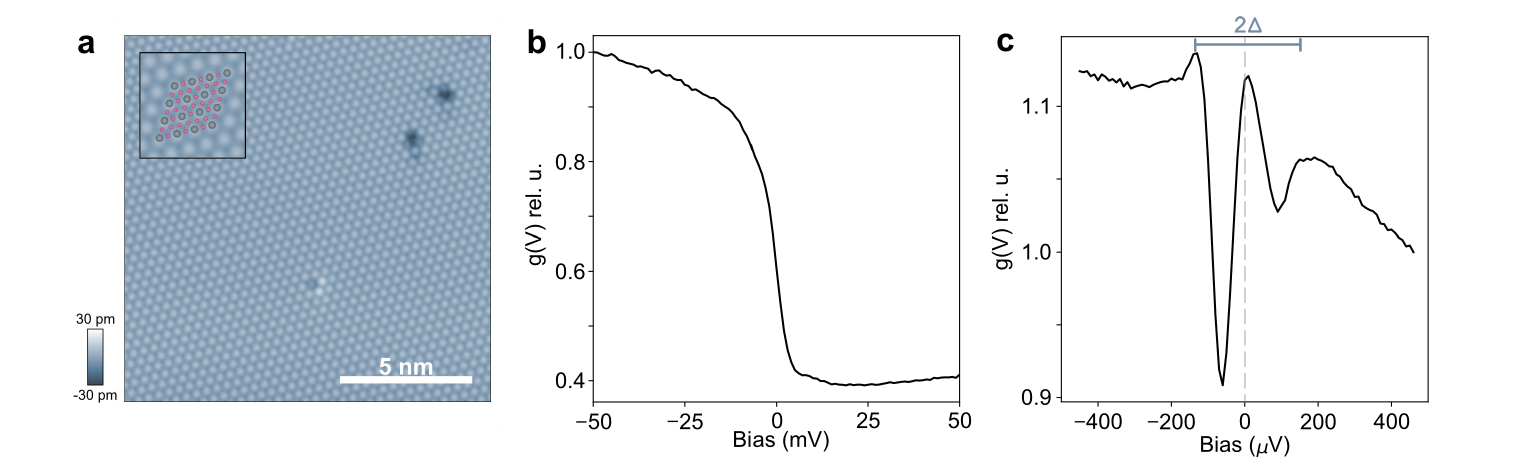}
    \caption{{\bf Imaging and spectroscopy of \ce{UPt3}.} (a) Topography of the surface of the \ce{UPt3} crystal ($T = 10\mathrm{K}$, $V_\mathrm{s} = 100\mathrm{mV}$, $I_\mathrm{s} = 300 \mathrm{pA}$). The inset shows a close up with the crystal structure superimposed. (b) Tunneling spectrum, $g(V)$, recorded in the normal state, showing a huge asymmetry of the density of states across the Fermi energy ($T = 10\mathrm{K}$, $V_\mathrm{s} = 50\mathrm{mV}$, $I_\mathrm{s} = 400 \mathrm{pA}$, $V_\mathrm{mod} = 1\mathrm{mV}$). (c) Tunneling spectrum $g(V)$ recorded in the superconducting state  taken at $T = 40\mathrm{mK}$, showing a clear gap-like structure. The size of the of superconducting gap $2\Delta = 280 \mu V$ is indicated ($V_\mathrm{s} = 2\mathrm{mV}$, $I_\mathrm{s} = 800\mathrm{pA}$, $V_\mathrm{mod} = 10 
    \mathrm{\mu V}$).}
    \label{fig:topospec_fig}
\end{figure}

Despite the rather isotropic bond lengths in \ce{UPt3}, the samples cleave along the basal $(0001)$ plane resulting in atomically clean surfaces with only few defects (Fig.~\ref{fig:topospec_fig}(a)), with large terraces with a typical width of $\sim 100\mathrm{nm}$ and a defect concentration of one defect in every 50 unit cells. The surface morphology shows clear signatures of the uranium atoms. Normal state tunneling spectra reveal a strong particle-hole asymmetry with a large change in the differential conductance when going across the Fermi energy (Fig.~\ref{fig:topospec_fig}(b)), suggesting a band in the occupied states that terminates right at the Fermi energy. The particle-hole asymmetry in the normal state density of states results in a change of the differential conductance by about $50\%$ when going from negative to positive bias within $50\mathrm{meV}$, and even on the scale of $1\mathrm{meV}$ the normal state differential conductance changes by more than $10\%$. The spectrum reveals little variation across the surface.

Tunneling spectra acquired at temperatures below $T_\mathrm{c}\approx 500\mathrm{mK}$ reveal a distinct structure at the Fermi energy. Fig.~\ref{fig:topospec_fig}(c) shows a spectrum acquired at $40\mathrm{mK}$ which exhibits two minima and three peaks. Most notably, and unlike the structure of the superconducting gap of a conventional superconductor, it is not the deepest minimum, but the central peak which is pinned at the Fermi energy. The total width of the gap-like structure is only about $280\mathrm{\mu eV}$ from peak to peak.

%
\begin{figure}
    \includegraphics[width=1\linewidth]{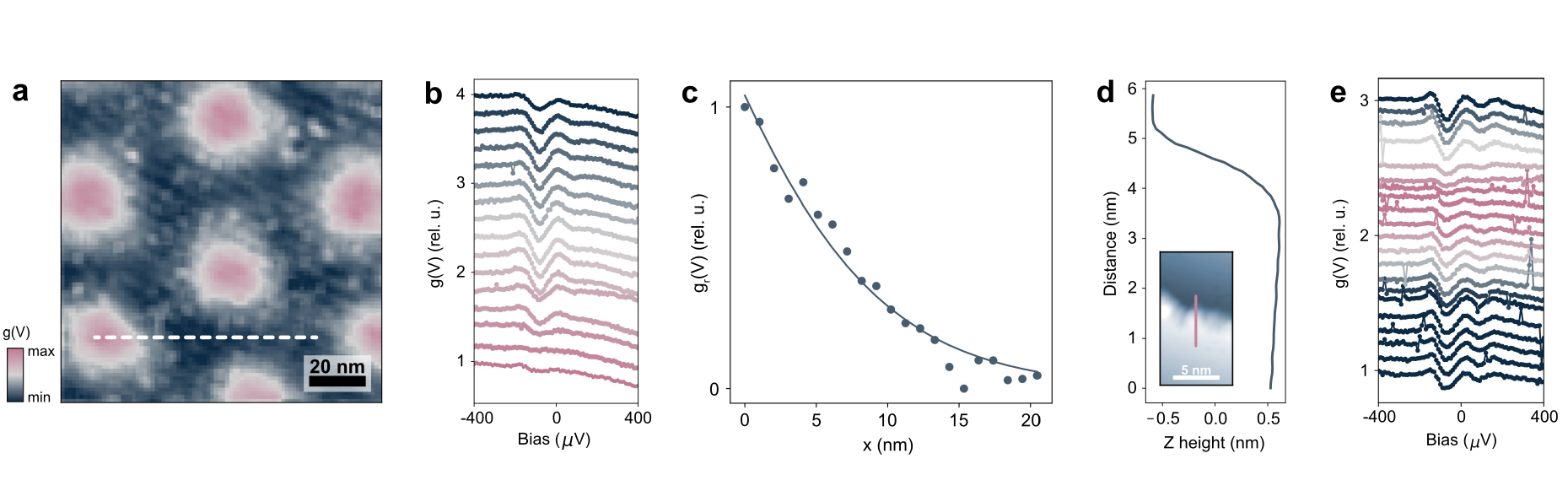}
    \caption{{\bf Vortex lattice and coherence length.} (a) Real space map of tunneling conductance, $g(V)$, for $V = -40\mathrm{\mu V}$ taken in a magnetic field $B = 0.75\mathrm{T}$ at $T = 40\mathrm{mK}$, showing the vortex lattice. In the vortex cores, $g(V)$ reverts to the normal state density of states ($V_\mathrm{s} = 2\mathrm{mV}$, $I_\mathrm{s} = 800\mathrm{pA}$, $V_\mathrm{mod} = 50 \mathrm{\mu V}$). $3\times3$ pixel averaging applied to an $86\times86$ pixel map. (b) $g(V)$ spectra taken from left to right along the white dashed line in (a), showing suppression of the gap within vortices ($V_\mathrm{s} = 2\mathrm{mV}$, $I_\mathrm{s} = 800 \mathrm{pA}$, $V_L = 10 \mathrm{\mu V}$). (c) Radially averaged $g_r(V)$, with a constant background value subtracted and then normalised, measured from the centre of a vortex core outwards. Data fit to $g_r(V)=1-\tanh(r/\xi)$ where $\xi=126 $Å, shown by the solid line. (d) Linecut of the step edge along the pink line in the inset; inset is a topography of a step edge ($V_\mathrm{s} = 10 \mathrm{mV}$, $I_\mathrm{s} = 400\mathrm{pA}$), showing a step height of 1.3 nm. Tunneling spectra along the pink line (e) show suppression of the gap at the step edge ($T = 40 \mathrm{mK}$, $V_\mathrm{s} = 2\mathrm{mV}$, $I_\mathrm{s} = 800 \mathrm{pA}$, $V_L = 10 \mathrm{\mu V}$).}
    \label{vortexlattice}
\end{figure}

To prove the origin of the gap-like structure, we have applied a small magnetic field along the surface normal. A superconductor will try to expel the field through screening currents, in the case of a type-II superconductor through the formation of a vortex lattice. Within the cores of these vortices, the superconductivity is expected to be suppressed. In Fig.~\ref{vortexlattice}(a), we show a spatial map of the differential conductance at a bias voltage of $-40 \mathrm{\mu V}$ in a field of $\mu_0 H= 0.75\mathrm{T}$. One can clearly see seven vortex cores arranged in a hexagonal lattice, as would be expected for Abrikosov vortices. The number of vortices is consistent with flux quanta of $\Phi_0=\frac{h}{2e}$. $g(V)$ spectra acquired in the vortex cores are indeed flat, confirming the suppression of superconductivity in the vortex cores and firmly establishing the feature in the spectra as the superconducting gap of \ce{UPt3}. From the length scale over which the gap recovers, we can estimate a coherence length $\xi\sim 126\mathrm{\AA}$.
We similarly observe a clear suppression of the superconducting gap near step edges and some of the point defects. Traversing the tunneling spectra across a step edge (Fig.~\ref{vortexlattice}(d, e)) shows a suppression of the superconducting gap right at the step edge, recovering on a similar length scale as close to a vortex core when moving away from the step edge.

\begin{figure}
    \includegraphics[width=1\linewidth]{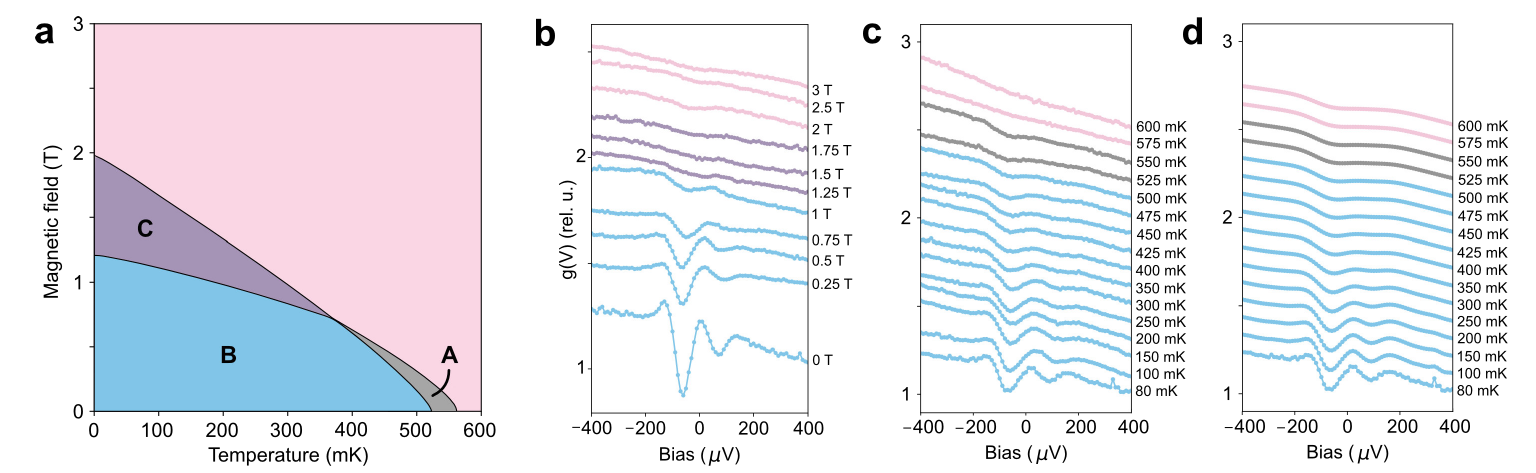}
    \caption{{\bf Temperature and field dependence of the gap.} (a) Low temperature phase diagram of \ce{UPt3}, with $\mu_0 H \parallel c$ showing the three superconducting phases (based on \cite{Adenwalla1990, Suderow1997, Huxley2000}). (b) Magnetic field dependence ($B \parallel c$) of $g(V)$ spectra recorded at $T = 45 \mathrm{mK}$. The gap is completely suppressed at fields larger than $
    \mu_0 H=1\mathrm{T}$ ($V_\mathrm{s} = 2\mathrm{mV}$, $I_\mathrm{s} = 800\mathrm{pA}$, $V_\mathrm{mod} = 10 \mathrm{\mu V}$). (c) Temperature dependence of $g(V)$ spectra, showing that the gap structure disappears above $550\mathrm{mK}$ ($V_\mathrm{s} = 2\mathrm{mV}$, $I_\mathrm{s} = 800\mathrm{pA}$, $V_\mathrm{mod} = 10 \mathrm{\mu V}$). (d) Tunneling spectrum taken at the lowest temperature but with simulated thermal broadening, but otherwise constant shape of the gap. The simulation shows that indeed in the experiment the gap is suppressed at $T=550
    \mathrm{mK}$, and does not just become indiscernable. The blue, purple, grey and pink colours in b-d indicate the B-phase, C-phase, A-phase and normal state as in panel a, respectively.}
    \label{fig:SC_state_fig}
\end{figure}

One of the most intriguing properties of the superconductivity in \ce{UPt3} is the phase diagram as a function of temperature and magnetic field (Fig.~\ref{fig:SC_state_fig}(a)), which from a range of macroscopic measurements suggests the existence of three separate superconducting phases in the material \cite{Adenwalla1990,hasselbach_superconducting_1990}. These phases vary somewhat dependent on the quality of the samples and the method used to extract them, but have been reported by multiple groups \cite{joynt_superconducting_2002}. To relate our results to the superconducting phase diagram, we have explored the magnetic field and temperature dependence of the gap. Fig.~\ref{fig:SC_state_fig}(b) shows the magnetic field dependence for fields between $0\mathrm{T}$ up to $3\mathrm{T}$. Our spectra show a slow suppression of the gap with increasing field, and a complete suppression between $1\mathrm{T}$ and $1.25\mathrm{T}$, with no further significant feature for fields $B\ge1.25\mathrm T$. This suggests that we observe the gap in the B-phase, but not in the C-phase.
The temperature dependence reveals a similar picture: with increasing temperature, the gap-like structure becomes weaker until the spectrum becomes flat at temperatures larger than $550\mathrm{mK}$. There is no evidence for a second transition in the temperature dependence. The temperature dependence is clearly stronger than what would be observed if thermal broadening was the only factor in suppressing it, Fig.~\ref{fig:SC_state_fig}(d), confirming a mean-field behaviour of the gap.

\section{Discussion}
\ce{UPt3} has long been a leading candidate for spin-triplet superconductivity: it was suggested that the phase diagram which is reminiscent of that of $^3\mathrm{He}$ could only be explained with an order parameter with triplet pairing \cite{Sauls_order_1994}. $\mu$SR \cite{luke_muon_1993} measurements suggested a time reversal symmetry breaking order parameter, which however was not confirmed in higher-quality samples \cite{de_reotier_absence_1995}. 
Measurements on Josephson junctions in different directions and close to $T_\mathrm c$ suggest a node in the order parameter in the A-phase \cite{Strand2009}.
Later theoretical work showed that the phase diagram can also be accounted for with a singlet order parameter \cite{Park_E1g_1996}. Our results demonstrate detection of the superconducting gap and a surface Andreev bound state in \ce{UPt3}. The size of the gap on the order of $2\Delta\sim280\mathrm{\mu eV}$ yields a gap to $T_\mathrm c$ ratio $\frac{2\Delta}{k_\mathrm BT_\mathrm c}\sim 5.9$, putting \ce{UPt3} in the strong coupling regime, with a $2\Delta/k_\mathrm BT_\mathrm c$ ratio remarkably close to that in other heavy fermion superconductors \cite{allan_imaging_2013,zhou_visualizing_2013,enayat_superconducting_2016}. Also in line with how the gap appears in STM measurements of other heavy fermion materials \cite{allan_imaging_2013,zhou_visualizing_2013,enayat_superconducting_2016,jiao_chiral_2020,herrera_quantum-well_2023}, the gap is rather shallow, with a reduction in differential conductance by only about $15\%$, suggesting that the surface might only be superconducting by proximity effect. 
Our measurements provide clear benchmark results for the symmetry of the order parameter: they put constraints on its out-of-plane sign structure and therefore on the overall pairing symmetry. Our measurements pick up a structure that can be directly associated with the superconducting gap, and exhibits a pronounced Andreev bound state at zero bias. 

To evaluate the order parameter, we have performed extensive theoretical modelling of the Andreev bound states for the different candidate symmetries. For this, we start from the Bogoljubov-de-Gennes Hamiltonian in the Nambu spinor basis and introduce the gap functions $\hat{\Delta}(\mathbf{R})$ to describe the superconducting pairing. For triplet order parameters, $\hat{\Delta}(\mathbf{R})$ takes the form
\begin{equation}
    \hat{\Delta}(\mathbf{R}) = \Delta\cdot\Gamma(\mathbf{R})\cdot ((\mathbf{d}\cdot\hat{\bm{\sigma}})\cdot i\hat{\sigma}_y) = \Delta\cdot\Gamma(\mathbf{R})\cdot\begin{pmatrix}
        -d_x +i\cdot d_y &d_z\\ d_z & d_x +i\cdot d_y
    \end{pmatrix},
\end{equation}
where the $\mathbf{d}$ vector encodes the direction of the spin of the Cooper pair, $\Gamma(\mathbf{R})$ is the spatial part of the wave function and $\Delta$ the pairing strength. We then calculate the surface density of states using an iterative surface Green's function method (see materials and methods, supplementary section \ref{tbdossection} for details). The calculations reveal that both triplet ($A_{2u}$, $E_{2u}$) and singlet ($E_{1g}$) order parameters result in Andreev surface bound states, however for the triplet order parameters, the Andreev bound state is not robust against Rashba spin splitting (see suppl. Figs. \ref{fig:RashbaDependence},~\ref{fig:DOS_plots}) \cite{kobayashi_fragile_2015}. For a material consisting of comparatively heavy elements such as Pt and U, spin-orbit coupling is important and significantly affects the bulk electronic structure. At the surface, the spin-orbit coupling results in sizeable Rashba spin-splitting, which we also find in Density Functional Theory (DFT) calculations (see suppl. Fig.~\ref{rashba}, suppl. sect. \ref{rashbasection}). Once a finite Rashba spin splitting, $\lambda \left(\hat{\sigma}_xk_y-\hat{\sigma}_yk_x\right)$, is accounted for, the only order parameter consistent with our experiments is the singlet $E_{1g}$ order parameter with a chiral spatial structure when accounting for the symmetry of the material. 
This is also supported by our observation of a vortex lattice which is fully consistent with an Abrikosov lattice. The density of vortices and the absence of zero bias peaks in the vortex cores suggest that they all contain one flux quantum. We note that this leaves open the possibility of the formation of unconventional vortex cores at the boundaries of chiral domains \cite{garcia-campos_visualization_2025}.

At the surface of the material, we appear to be mostly sensitive to the B phase, with no indication of a double transition either as a function of field or temperature. As a function of field, we do not observe a gap in the region that is ascribed to the C phase, whereas as a function of temperature, we observe only the disappearance of the gap at the transition to the normal state. It might be that the A and C phases have no characteristic gap features allowing to distinguish them in tunneling spectroscopy. However, it is also possible that the A and C phases simply do not exist at the surface of the material. Indeed, it has been suggested that the phase diagram might be a consequence of superconductivity nucleating at defects or stacking faults \cite{hayden_afmorder_1992, brison_magnetism_1994,joynt_superconducting_2002}.

From our measurements, a number of new questions arise: (1) the symmetry of the order parameter suggests an unconventional pairing mechanism which should be captured by random phase approximation (RPA) or similar calculations of the pairing vertex. While some work has been done, focusing on triplet instabilities \cite{nomoto_exotic_2016}, a more in-depth comparison with such calculations would be highly valuable to assess the fidelity of the theories for spin-fluctuation mediated superconductivity for uranium-based superconductors. (2) so far, no evidence of a neutron resonance that would support a sign-changing order parameter has been reported. Our results constrain the location in $\mathbf{q}$-space where such such a neutron resonance is expected, and its detection would provide important additional evidence for the symmetry. (3) The absence of signatures of the transitions to the A and the C phases in our measurements suggests that a re-evaluation of the superconducting phase diagram is warranted -- if indeed nucleation of superconducting phases at imperfections plays a role, new, more sensitive methods that can be applied to microcrystals promise a better understanding of the origin of the superconducting phases. (4) the strong asymmetry of the normal state density of states is notable and to the best of our knowledge not typically accounted for in theoretical descriptions of the gap structure. It suggests a band edge or Van Hove singularity very close to the Fermi energy, as inferred previously from quantum oscillations \cite{mccollam_lifshitz_2021}. The steep drop in density of states across the Fermi energy is bound to have implications for the superconducting properties.

\section{Methods}
\subsection{Crystal growth}
The crystal was grown by the Czochralski method from a stoichiometric mixture of depleted uranium and high-purity platinum under ultra-high vacuum in a RF cold crucible furnace and separately annealed for $950^\circ\mathrm{C}$. The samples were in addition annealed to $900^\circ$ for seven days.

\subsection{Scanning tunneling microscopy}
Measurements were performed using a home-built ultra-low temperature STM mounted in a dilution refrigerator \cite{singh_construction_2013}. The instrument routinely operates at temperatures well below $100\mathrm{mK}$ with a base temperature of $20\mathrm{mK}$ and in magnetic fields up to $14\mathrm{T}$. The bias voltage $V$ is applied to the sample, with the tip at virtual ground. Differential conductance spectra were acquired using a lock-in amplifier, adding a bias modulation to the sample bias and recording the response in the current.
\ce{UPt3} samples were mounted on a sample holder with a notch cut in the side and cleaved \textit{in-situ} at a cleaving stage mounted to the $4\mathrm{K}$ plate of the cryostat and following cleavage immediately transferred to the STM head.

\subsection{Density Functional Theory calculations}
Density functional theory calculations were performed using the projector augmented plane-wave method \cite{Blochl_PAW_1994}, as implemented in the Vienna Ab Initio Simulation Package \cite{Kresse_VASP_1993, Kresse_VASP_1996, Kresse_VASP_1999}. Calculations have been performed including relativistic corrections to account for spin-orbit interactions and allowing for non-collinear spin configurations. A slab with 40 atoms in the unit cell was used, composed of 5 primitive unit cells stacked in the $c$-direction, with lattice parameters $a = 5.764$ \r{A} and $c = 4.992$ \r{A} \cite{Heal_Structure_1955}, and a $20$ \r{A} vacuum above the surface. We used an $800\mathrm{ eV}$ plane-wave energy cut off and a $6 \times 6 \times 1$ Monkhorst-Pack $\mathbf{k}$-grid.

\subsection{Calculations of Andreev bound states}
We have performed surface Green's function calculations to model the Andreev bound states using the St Andrews calcQPI code \cite{wahl_calcqpi_2025,wahl_calcqpi_code_2025}. To describe the different superconducting order parameters, we used a nearest neighbour tight-binding model on a hexagonal lattice and added the pairing interaction in real space. We have included a Rashba spin-orbit coupling term for the surface layer, and then used an iterative Green's function scheme as implemented in calcQPI to obtain the surface Green's function \cite{guinea_effective_1983,sancho_quick_1984,sancho_highly_1985}. For details see supplementary material. 

\noindent{\bf Acknowledgements:} We acknowledge initial cleaving studies by Jean-Philippe Reid. RB gratefully acknowledges support through the EPSRC Doctoral Training grant EP/W524505/1. LCR and PW acknowledge funding from the Leverhulme Trust through Research Project Grant RPG-2022-315, PW from the Engineering and Physical Sciences Research Council through EP/X015556/1, and CM and PW through UKRI1107. HD gratefully acknowledges support through the Grant family scholarship programme. This work used computational resources of Archer2 in Edinburgh, the high-performance computing cluster Hypatia at the University of St Andrews, and the UK National Tier-2 HPC Service Cirrus at EPCC (http://www.cirrus.ac.uk) with project code sc148 funded by The University of Edinburgh, the Edinburgh and South East Scotland City Region Deal, and UKRI via EPSRC.\\

\noindent{\bf Author Contributions:} RB did STM measurements, with early feasibility studies by MN and AM. LCR, RB and PW did calculations of the Andreev bound states; HD and RB did DFT calculations supervised by CM. RB prepared the figures in the main manuscript. AH grew and characterized the samples. PW initiated and supervised the project. PW, RB and LCR wrote the manuscript with contributions from all authors. All authors discussed the manuscript.

\noindent{\bf Competing Interests:} The authors declare that they have no competing interests.\\

\bibliographystyle{unsrtnat}
\bibliography{upt3articles}


\newpage
\renewcommand{\thefigure}{S\arabic{figure}}
\renewcommand{\thetable}{S\arabic{table}}
\renewcommand{\theequation}{S\arabic{equation}}
\renewcommand{\thesection}{S\arabic{section}}
\setcounter{section}{0} 
\setcounter{figure}{0}
\setcounter{table}{0}
\setcounter{equation}{0}

\begin{center}
\large{Supplementary Material for 'Determining the superconducting order parameter of \ce{UPt3} using tunneling microscopy'}
\end{center}
\vspace{1cm}
\author{Rebecca Bisset}
\affiliation{SUPA, School of Physics and Astronomy, University of St Andrews, North Haugh, St Andrews, KY16 9SS, United Kingdom}
\author{Luke C. Rhodes}
\affiliation{SUPA, School of Physics and Astronomy, University of St Andrews, North Haugh, St Andrews, KY16 9SS, United Kingdom}
\author{Hugo Decitre}
\affiliation{SUPA, School of Physics and Astronomy, University of St Andrews, North Haugh, St Andrews, KY16 9SS, United Kingdom}
\author{Matthew J. Neat}
\affiliation{SUPA, School of Physics and Astronomy, University of St Andrews, North Haugh, St Andrews, KY16 9SS, United Kingdom}
\author{Ana Maldonado}
\affiliation{SUPA, School of Physics and Astronomy, University of St Andrews, North Haugh, St Andrews, KY16 9SS, United Kingdom}
\author{Andrew Huxley}
\affiliation{SUPA, School of Physics and Astronomy, University of Edinburgh, Kings Buildings, Edinburgh, EH9 3FD, United Kingdom}
\author{Carolina A. Marques}
\affiliation{SUPA, School of Physics and Astronomy, University of St Andrews, North Haugh, St Andrews, KY16 9SS, United Kingdom}
\author{Peter Wahl}
\affiliation{SUPA, School of Physics and Astronomy, University of St Andrews, North Haugh, St Andrews, KY16 9SS, United Kingdom}
\affiliation{Physikalisches Institut, Universität Bonn, Nussallee 12, 53115 Bonn, Germany}




\section{Modelling of Andreev bound states}
\label{andreevsection}
\subsection{Tight-binding model and density of states calculations}
\label{tbdossection}
\begin{figure}[b]
    \centering
    \includegraphics[width=0.99\linewidth]{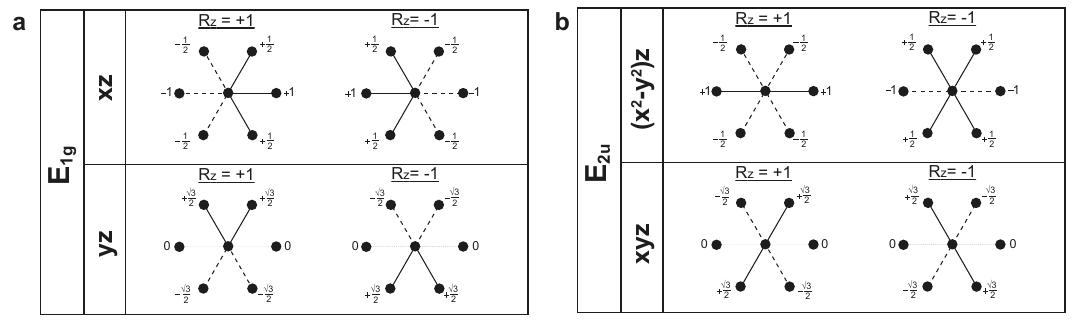}
    \caption{Lowest harmonic form factors for the real space superconducting order parameter on a hexagonal lattice with $D_{6h}$ point group symmetry for (a) an E$_{1g}$ order parameter with symmetry $xz+iyz$ and (b) an $E_{2u}$ order parameter with symmetry $(x^2-y^2)z + 2ixyz$. The solid (dashed) lines indicate the sign of the gap connecting nearest neighbour unit cells. }
    \label{fig:orderparameters}
\end{figure}
To calculate the surface density of states for the different candidates for the superconducting order parameters, we start from the Bogoliubov-de-Gennes Hamiltonian for a tight binding model in real space

\begin{equation}
\hat{H}_{s}=\begin{bmatrix}
           \hat{H_0}(\mathbf{R}) & \hat{\Delta}(\mathbf{R}) \\
           \hat{\Delta}^\dagger(\mathbf{R}) & -\hat{H_0^T}(-\mathbf{R})
         \end{bmatrix},
\label{Eq:bdg_Hamiltonian}
\end{equation}

\noindent where $\hat{H}^0(\mathbf{R})$ and $\hat{\Delta}(\mathbf{R})$ are $2\times 2$ block matrices in the spin basis, and $\mathbf{R}$ are vectors pointing to the nearest neighbours. We choose $\hat{H}^0(\mathbf{R})$ to be a three-dimensional nearest neighbour tight binding Hamiltonian on a hexagonal lattice with hopping $t=1\mathrm{eV}$, resulting in a band width of $12\mathrm{eV}$. 

\noindent For superconductors which exhibit spin-singlet pairing, the gap function that connects the particle and anti-particle sector can then be written as 

\begin{equation}
    \hat{\Delta}(\mathbf{R}) = \Delta\cdot\Gamma(\mathbf{R})\cdot i\hat{\sigma}_y = \begin{pmatrix}
        0 &\Delta\cdot \Gamma(\mathbf{R})\\-\Delta\cdot\Gamma(\mathbf{R}) & 0
    \end{pmatrix},
\end{equation}

\noindent where $\Gamma(\mathbf{R})$ is the symmetry function that describes the real space gap structure, and obeys $\Gamma(\mathbf{R}) = \Gamma(\mathbf{-R})$. For a spin triplet superconductor $\Gamma(\mathbf{R}) = -\Gamma(\mathbf{-R})$ and the superconducting gap takes the form
\begin{equation}
    \hat{\Delta}(\mathbf{R}) = \Delta\cdot\Gamma(\mathbf{R})\cdot ((\mathbf{d}\cdot\hat{\bm{\sigma}})\cdot i\hat{\sigma}_y) = \Delta\cdot\Gamma(\mathbf{R})\cdot\begin{pmatrix}
        -d_x +i\cdot d_y &d_z\\ d_z & d_x +i\cdot d_y
    \end{pmatrix},
\end{equation}

\noindent where $d_x$, $d_y$, $d_z$ denote the components of the $\mathbf{d}$-vector and $\Delta$ is the superconducting pairing strength. 

\noindent From the Hamiltonian in eq.~\ref{Eq:bdg_Hamiltonian} we then obtain the bulk and surface Green's function following an iterative Green's function scheme \cite{guinea_effective_1983,sancho_quick_1984,sancho_highly_1985} using the implementation in calcQPI \cite{wahl_calcqpi_2025,wahl_calcqpi_code_2025} to calculate the density of states. A 2D k-grid of $4096\times 4096$ points was employed, and an energy broadening of 200~$\mu$eV was added to the Green's function. 

\subsection{Rashba spin-orbit coupling}
\label{rashbasc}
For the surface layer, we add a Rashba-spin-orbit coupling term $H_\mathrm{soc}$ \cite{bychkov_properties_1984} in real space of the form 
\begin{equation}
    \hat{H}_\mathrm{soc}(\mathbf{R}) = \lambda \mathbf{e}_z \cdot \left(\hat{\sigma}\times\mathbf{R}_\textrm{NN}\right)
\end{equation}
to the nearest-neighbour hopping terms of the Hamiltonian of the surface layer $\hat{H}_0^\mathrm{s}$, where the vectors $\mathbf{R}_\textrm{NN}$ are to the nearest neighbour atoms and $\mathbf{e}_z$ is a unit vector along the surface normal. As above, the vectors $\mathbf{R}_\textrm{NN}$ are relative to the lattice vectors, so are dimensionless.    
\noindent In Fig.~\ref{fig:intro_fig}(b, c) of the main text we use values $t=1$~eV as well as a rigid chemical potential shift of $\mu=300\mathrm{meV}$ to remove the van Hove singularity from the Fermi level. We set $\lambda = t/2 = 500\mathrm{meV}$ and $\Delta=t/100 = 10\mathrm{meV}$. We consider superconducting pairing with a $\mathbf{d}$-vector $\mathbf{d}\|\mathbf{e}_z$. 

\begin{figure}
    \centering
    \includegraphics[width=0.7\linewidth]{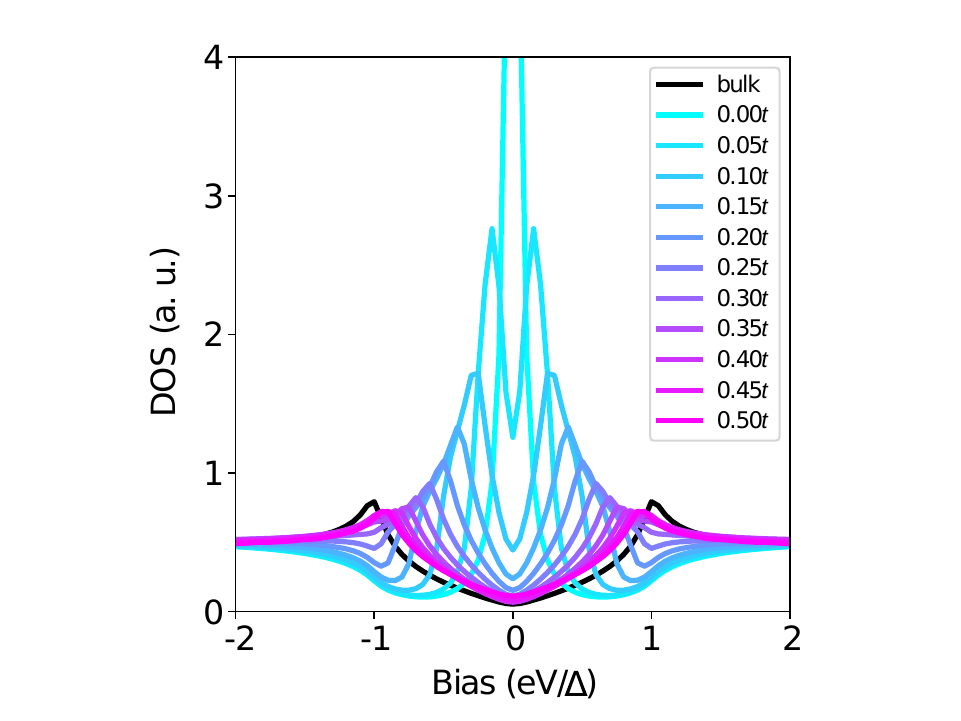}
    \caption{Surface density of states for an $A_{2u}$ order parameter as a function of Rashba spin-orbit coupling $\lambda$. Even for a Rashba spin splitting as small as $\lambda=0.05t$ where $t$ is the nearest neighbour hopping, a finite splitting of the zero bias Andreev bound state can be observed. This splitting increases with increasing Rashba spin splitting, until it merges with the bulk superconducting coherence peaks.}
    \label{fig:RashbaDependence}
\end{figure}

In Fig.~\ref{fig:RashbaDependence}, we plot the surface DOS as a function of Rashba spin splitting for the $A_{2u}$ triplet order parameter. Only the ratio $\lambda/t$ influences this splitting, and we note that a splitting as small as $0.05t$ splits the zero bias Andreev bound state substantially. Hence, for a system with a dominant nearest-neighbour hopping value of 1eV (and thus a bandwidth of 12~eV for a hexagonal system) the splitting would be observable for a Rashba spin orbit interaction on the order of 50~meV. For smaller dominant hopping, this would also scale to a smaller value. 

We note that the suppression of the bound state depends on the $\mathbf{d}$-vector, for a chiral $\mathbf{d}$-vector, the bound state would not be suppressed, however that is not an order parameter proposed for \ce{UPt3}.

\subsection{Superconducting order parameters}
\begin{figure}
    \centering
    \includegraphics[width=0.99\linewidth]{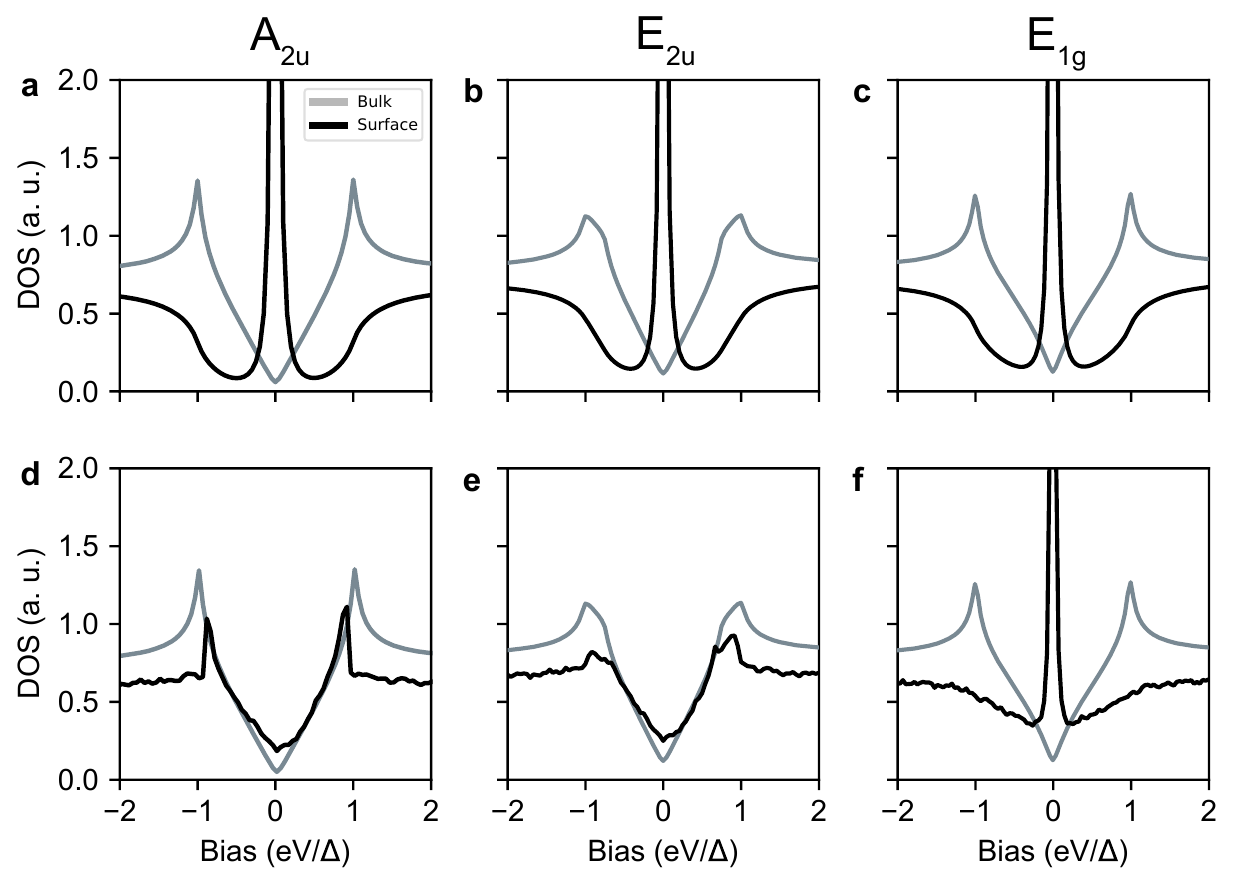}
    \caption{DOS plots calculated for the bulk, in grey, and surface, in black, both without Rashba spin-orbit coupling (panels (a)-(c)) and with (panels (d)-(f)), for the \ce{A_{2u}}, \ce{E_{2u}} and \ce{E_{1g}} order parameters, respectively.}
    \label{fig:DOS_plots}
\end{figure}

Considering the $D_{6h}$ point group symmetry of the lattice of \ce{UPt3}, we can write the form factors, $\Gamma(\mathbf{R})$ that conform to the $E_{1g}$ and $E_{2u}$ irreducible representations as shown in Fig.~\ref{fig:orderparameters}. These symmetries are both doubly degenerate due to their chirality and can be written as $E_{1g} = (xz+iyz)$ and $E_{2u}=(x^2-y^2)z + 2ixyz$, respectively. We obtain the parameters for the tight-binding model from these functions by evaluating them along the nearest neighbour directions to obtain the values for each $\mathbf{R}$-vector on the hexagonal lattice.

For Andreev bound states to form for a surface in the $x$-$y$-plane, the superconducting order parameter must satisfy $\Delta(R_\parallel,R_z) = -\Delta(R_\parallel,-R_z)$. From analysis of the  character table of the $D_{6h}$ point group, this is only satisfied for three symmetries, the $E_{1g}$ and $E_{2u}$ order parameters, and the $A_{2u}$ order parameter \cite{kobayashi_fragile_2015}, with a lowest-harmonic symmetry function defined by $A_{2u} = z$. For completeness, in Fig.~\ref{fig:DOS_plots} we plot the bulk and surface density of states for all three symmetries with and without Rashba spin-orbit coupling. For the two triplet order parameters, $A_{2u}$ and $E_{2u}$, the finite Rashba term suppresses the zero energy Andreev bound state. 

We note that also for an $s$-wave order parameter and a magnetic surface one would also obtain an Andreev bound state at the surface, which would however again be split by a Rashba spin-orbit coupling. Therefore, the only case of a zero energy Andreev bound state that is robust against Rashba spin-splitting for the $D_{6h}$ point group that is relevant for \ce{UPt3} and a surface normal to the $c$-axis is for an $E_{1g}$ order parameter with singlet pairing.

\section{Magnitude of Rashba spin splitting at the surface of \ce{UPt3}}
\label{rashbasection}
\begin{figure}
    \includegraphics[width=0.99\linewidth]{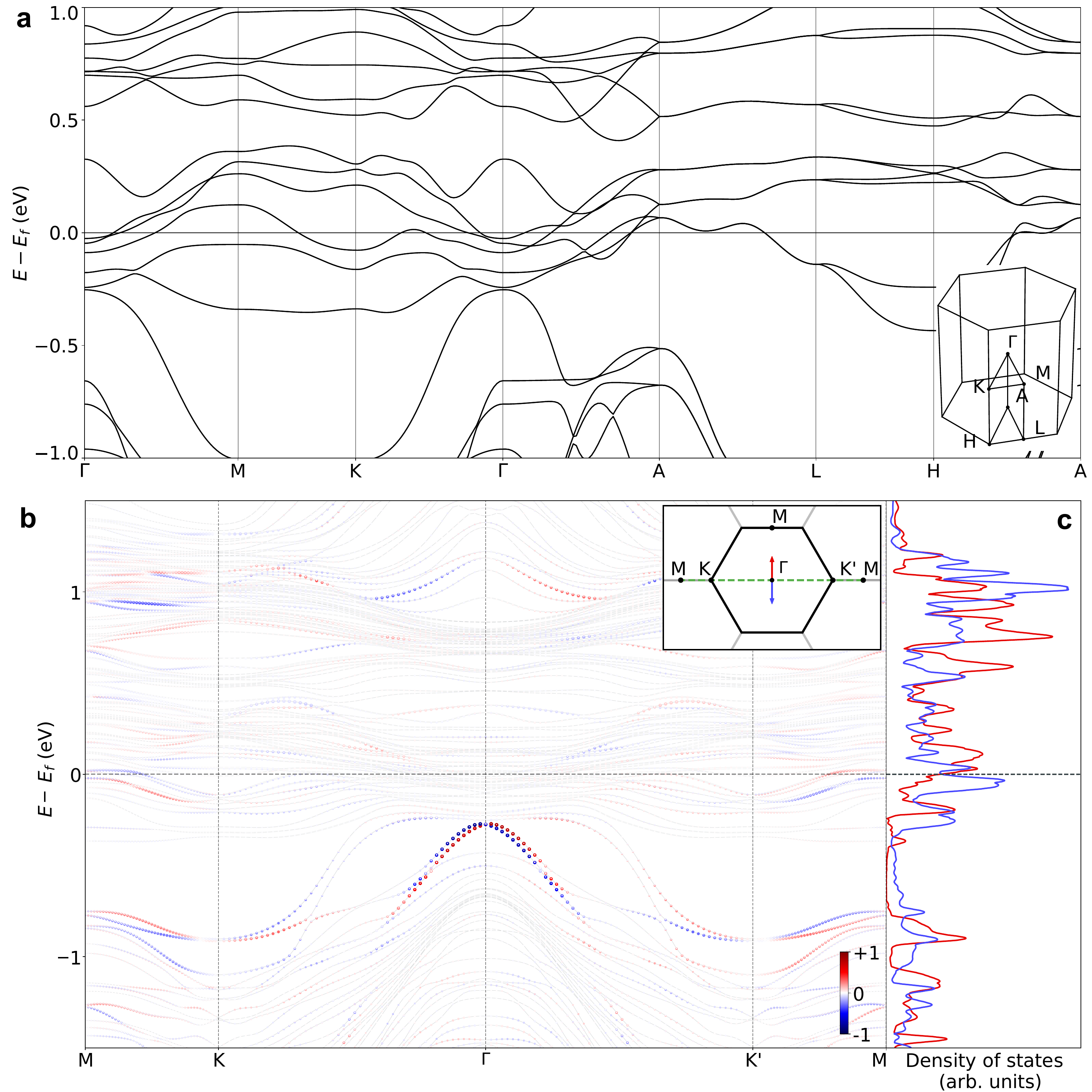}
    \caption{Rashba-spin splitting in \ce{UPt3}. (a) Bulk electronic structure along the path shown in the inset obtained from a DFT calculation with spin-orbit coupling. (b) Surface-projected and spin-resolved electronic structure for the path shown in green in the inset, showing multiple spin-split bands, notably for a surface state below $-0.2\mathrm{eV}$ as well as bands close to the Fermi energy in the K-M cut. The spin-projections are illustrated in the inset. (c) Spin-resolved and surface-projected density of states integrated along $\Gamma-K'-M$ showing clear spin-polarization of the states in the surface layer.}
    \label{rashba}
\end{figure}
The presence of a surface introduces inversion symmetry breaking which results in Rashba spin splitting of the bands near the surface. The magnitude of the Rashba spin splitting depends on the atomic spin-orbit coupling and details of the vacuum potential. The atomic spin-orbit coupling of uranium and platinum is non-negligible, suggesting a sizable Rashba spin splitting. To confirm this expectation and provide a quantitative assessment, we have performed Density Functional Theory (DFT) calculations for slabs of \ce{UPt3} including spin-orbit coupling. For the bulk, our calculations show a band structure that is consistent with previous DFT calculations where spin-orbit coupling has been accounted for\cite{nomoto_exotic_2016,mccollam_lifshitz_2021} and with quantum oscillations\cite{mccollam_lifshitz_2021}. The calculations show significant Rashba spin splitting in the band structure (compare fig.~\ref{rashba}), including for bands right at the Fermi energy. For some bands, the momentum splitting reaches comparable magnitude to the size of the Brillouin zone.

Our DFT calculations also provide an estimate of the band width. The band manifold that dominates around the Fermi energy has a band width significantly smaller than $1\mathrm{eV}$, closer to about $0.5\mathrm{eV}$. This provides an upper bound to the bandwidth, as correlation effects tend to reduce the bandwidth compared to the bare-band dispersion due to renormalization. Using $1\mathrm{eV}$ as an upper bound and the estimates from section~\ref{rashbasc} suggests that a Rashba spin splitting on the order of $5\mathrm{meV}$ would be sufficient to split the Andreev bound state in this system, significantly smaller than the spin splittings seen in the calculation. We note that it has previously been shown that correlation effects tend to increase spin-orbit coupling.\cite{tamai_high-resolution_2019}


\end{document}